\documentclass[11pt]{article}

\usepackage[labelfont=bf]{caption} % Bold figure number
\usepackage[font={small}]{caption}   %% FIGURE CAPTION FONT MODIFICATION

\usepackage{graphicx}% Include figure files
\usepackage{dcolumn}% Align table columns on decimal point
\usepackage{bm}% bold math
\usepackage{hyperref}
\usepackage[utf8]{inputenc}
\usepackage{lipsum}  
\usepackage{chngpage}
\usepackage{amsmath}
\usepackage{authblk}
\usepackage{amsfonts}
\usepackage{soul}
\usepackage{color}%
\usepackage{placeins}
\usepackage{comment}
\usepackage[numbers, sort&compress]{natbib}  %%% For compressing the reference numbering
\usepackage{lineno}
\begin{document}

\title{Observation of trapped fractional charge and topological states at disclination defects in higher-order topological insulators}

\author[1]{Christopher W. Peterson}
\author[2]{Tianhe Li}
\author[1]{Wentao Jiang}
\author[2]{\\ Taylor L. Hughes}
\author[3]{Gaurav Bahl}
\affil[1]{\footnotesize{Department of Electrical and Computer Engineering, University of Illinois at Urbana-Champaign, Urbana, IL, USA}}
\affil[2]{\footnotesize{Department of Physics and Institute for Condensed Matter Theory, University of Illinois at Urbana-Champaign, Urbana, IL, USA}}
\affil[3]{\footnotesize{Department of Mechanical Science and Engineering, University of Illinois at Urbana-Champaign, Urbana, IL, USA}}

\date{\today}

\maketitle

%\linenumbers

%
\textbf{
Topological crystalline insulators (TCIs) can exhibit unique, quantized electric phenomena such as fractional electric polarization and boundary-localized fractional charge \cite{ZakBerry1989, VanderbiltElectric1993, King-SmithTheory1993, RestaMacroscopic1994, hughes2011inversion, TurnerQuantized2012}.
This quantized fractional charge is the generic observable for identification of TCIs that lack robust spectral features \cite{inversiontaylor,TurnerQuantized2012,miert2017}, including ones having higher-order topology \cite{QuadTheory, benalcazar2017electric, CornerCharge, FCA}.
It has been predicted that fractional charges can also manifest where crystallographic defects disrupt the lattice structure of TCIs, potentially providing a bulk probe of crystalline topology \cite{MiertDislocation2018,CornerCharge,LiuShift2019,tianheDisclin}. 
However, this capability has not yet been confirmed in experiment since measurements of charge distributions in TCIs have not been accessible until recently \cite{FCA}.
Here, we experimentally demonstrate that disclination defects can robustly trap fractional charges in TCI metamaterials, and show that this trapped charge can indicate non-trivial higher-order crystalline topology even in the absence of any spectral signatures.
Furthermore, we uncover a connection between the trapped charge and the existence of topological bound states localized at these defects.
We test the robustness of these topological features when the protective crystalline symmetry is broken, and find that a single robust bound state can be localized at each disclination alongside the fractional charge.
Our results conclusively show that disclination defects in TCIs can robustly trap fractional charges as well as topological bound states, and moreover demonstrate the primacy of fractional charge as a probe of crystalline topology.
}

\vspace{12 pt}

\begin{figure}
	\begin{adjustwidth}{-1in}{-1in}
    \centering
    \includegraphics[width = \linewidth]{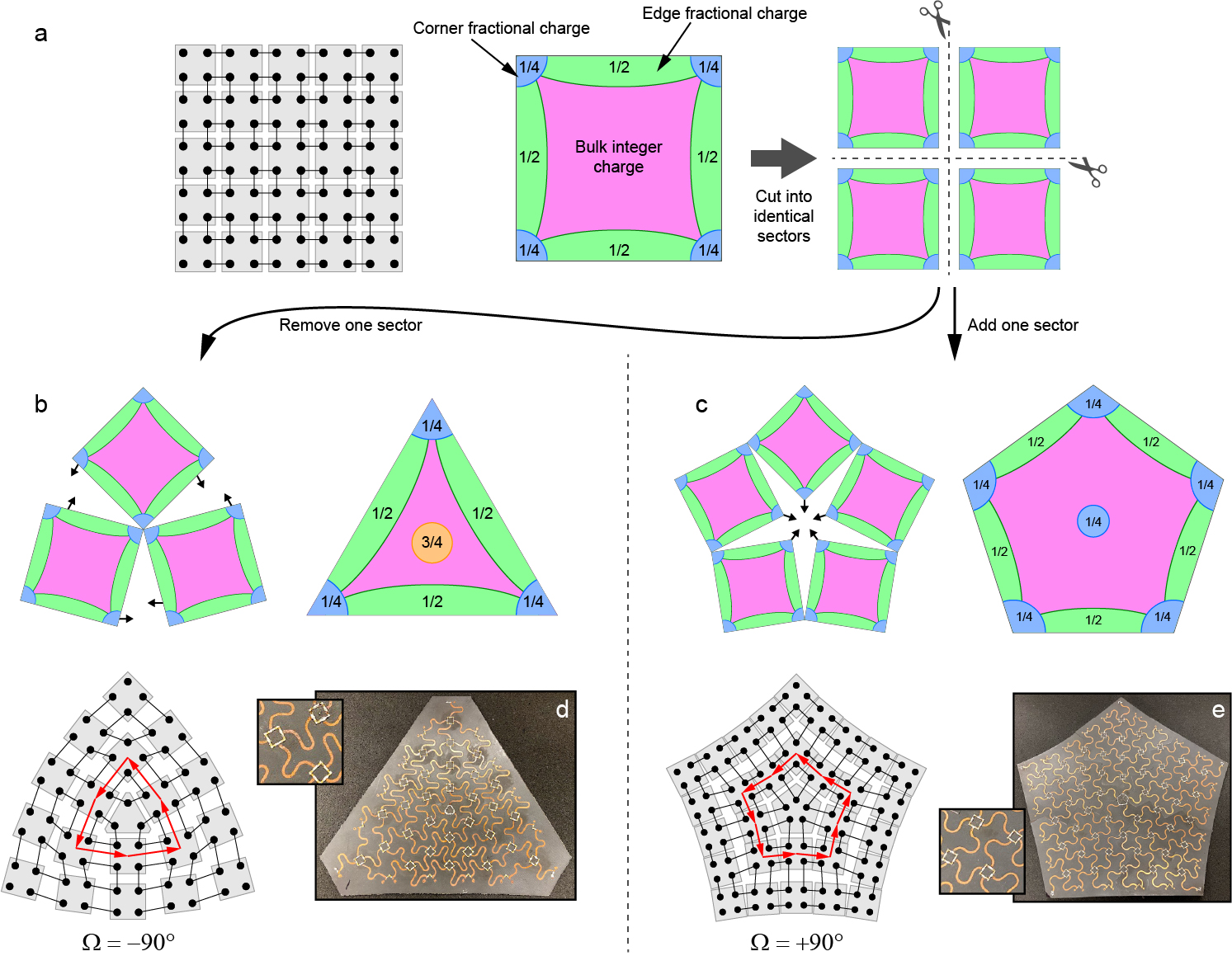}
    \caption{Fractional charge at disclination defects. (a) $C_4$-symmetric HOTI with $\frac{1}{4}$ corner charge and $\frac{1}{2}$ edge charge. A schematic of the tight-binding model (black dots are atoms, black lines are strong bonds) and illustration of the fractional charge are shown. Due to its $C_4$ symmetry, the HOTI can be cut into 4 identical sectors.
    (b) One sector is removed and the remaining three are glued together to create a disclination defect with a negative Frank angle $\Omega = -90^\circ$. Red arrows on the schematic show calculation of the Frank angle (see \ref{disclination}). The disclination traps a $\frac{3}{4}$ fractional charge, since it is formed by combining three $\frac{1}{4}$ charge corners.
    (c) One sector is added and the resulting five are glued together to create a disclination defect with a positive Frank angle $\Omega = +90^\circ$. Red arrows on the schematic show calculation of the Frank angle (see \ref{disclination}). The disclination traps a $\frac{1}{4}$ fractional charge, since it is formed by combining five $\frac{1}{4}$ charge corners.
    (d) Photo of fabricated microwave-frequency circuit TI with disclination having Frank angle $\Omega=-90^\circ$ (e) Photo of fabricated circuit with disclination having Frank angle $\Omega=+90^\circ$. Insets show $C_4$-symmetric bulk coupling.}
    \label{fig1}
    \end{adjustwidth}
\end{figure}

Topological insulators (TIs) are materials characterized by quantized topological invariants that are defined with respect to symmetries of their gapped bulk Hamiltonian \cite{schnyder2008classification3D,schnyder2008classification,hasan2010colloquium,qi2011topological}.
Topological invariants remain fixed as long as the symmetries and the bulk bandgap are preserved, which grants TIs remarkably robust properties that can survive even in the presence of strong disorder \cite{qi2011topological,chiu2016classification,schnyder2008classification}.
The most well-known class of TIs are the strong topological insulators, which are protected by local symmetries (time reversal symmetry, particle-hole symmetry, and/or chiral symmetry) and host robust gapless states at boundaries one dimension lower than the bulk, e.g. edges in 2D and surfaces in 3D \cite{KaneQauntum2005, bernevig2006quantum, FuTopological2007, MooreTopological2007}.
These gapless boundary states have been essential in the experimental identification of TIs since they produce a clear and robust spectral signature within the bulk bandgap.
In addition to local symmetries, crystalline symmetries, e.g. mirror or rotation symmetry, can likewise protect topological invariants such as a quantized electric polarization \cite{fu2007topological, TCIsreview, TeoSurface2008, FuTopological2011, FuTopological2012, SlagerSpace2013, ShiozakiTopology2014, FangNew2015, WatanabeTopological2017}.
However, the topological crystalline insulators (TCIs) \cite{fu2007topological} protected by these symmetries may not always manifest spectral features in their bulk bandgap, as crystalline symmetries in many cases protect the degeneracy of boundary-localized states, but do not restrict their energy \cite{CornerCharge, FCA}.
Instead, these classes of TCIs can be identified by, e.g., the quantized fractional charge that manifests at their boundaries when crystalline symmetries quantize their bulk electric polarization in fractions of the electron charge $e$ \cite{SuSolitons1979, ZakBerry1989, King-SmithTheory1993, VanderbiltElectric1993, RestaMacroscopic1994, hughes2011inversion, fang2012bulk, TurnerQuantized2012}.
Certain higher-order TCIs having a vanishing polarization (i.e. vanishing dipole moment) are further characterized by higher electric multipole moments, such as a quantized quadrupole moment \cite{QuadTheory, benalcazar2017electric, Peterson2018, HuberQuadrupoleNature2018, ImhofTopolectrical2018, Mittal2019}.
These multipole moments manifest as quantized fractional charges at boundaries with higher co-dimensions, e.g. both a quadrupole moment in 2D and an octopole moment in 3D manifest fractional corner charge \cite{QuadTheory}.
Furthermore, it was recently shown that even some higher-order TCIs without a bulk multipole moment can be identified by anomalous fractional charge at their corners \cite{van2018higher,CornerCharge,FCA}.
Crystallographic defects such as dislocations or disclinations, which respectively break lattice translation and rotation symmetry, are expected to also trap fractional charges \cite{deJuan,MiertDislocation2018,ObispoFractional2014,CornerCharge,LiuShift2019,tianheDisclin}. 
This trapped fractional charge thus serves as the generic bulk probe of crystalline topology, enabling crystalline insulators to be characterized away from their boundaries.
In rotationally-symmetric TCIs, the fractional charge $Q$ (in units of $e$) trapped by a disclination generically satisfies \cite{TeoDisclin,BenalcDisclin,CornerCharge,tianheDisclin}
\begin{equation} \label{disc_index}
    Q = \frac{\Omega}{2\pi} \eta + \sum_{i,j=1,2} \epsilon_{ij} B_i P_j \mod 1,
\end{equation} 
where the Frank angle $\Omega$ and the Burgers vector ${\bf B}$ are respectively the amount of rotation and translation accumulated by a vector under parallel transport along a loop enclosing the disclination core, $\mathbf{P}$ is the electric polarization, and $\eta$ is an integer determined by the Wannier representation of the filled bulk bands ($\epsilon_{ij}$ is the Levi-Civita symbol, see \ref{disclination} for a detailed explanation of these quantities).
From Eqn.~\eqref{disc_index}, we find that the expected trapped charge depends on both the shape of the defect (through $\Omega$ and $\mathbf{B}$), and the band structure of the insulator (through $\eta$ and $\mathbf{P}$).
However, despite theoretical analyses predicting quantized fractional charge to manifest at defects in TCIs, measurements of the charge distribution within insulators have not been previously accessible, and experimental confirmation of this relation has remained elusive.
Recently, an equivalent of boundary-localized fractional charge has been measured experimentally in TCI metamaterials \cite{FCA}.
These metamaterials are neutral bosonic systems and do not carry literal electric charge, instead reflection spectroscopy can be used to measure their local density of states (DOS) directly with high spatial resolution.
By integrating the measured DOS over the frequency range of an entire bulk band, and normalizing to correctly count the number of modes in each unit cell, an analogous quantity to the charge density can be determined as if each state of the bulk band were filled by a single electron (see Methods for details).
This analogous quantity physically represents the density of modes, or the \emph{mode density}, in a unit cell over a given frequency range. 
Previous work has shown that measurements of mode density successfully capture previously inaccessible observables such as fractional corner charge, and can be used to identify both first-order and higher-order topology in TCIs \cite{FCA}.
Using this measurement method, here we report the experimental observation of fractional mode density trapped by two different disclination defects in a $C_4$-symmetric HOTI metamaterial: one with a negative Frank angle $\Omega = -90^\circ$, and one with a positive Frank angle $\Omega = +90^\circ$, as shown in Fig.~\ref{fig1}.
We find that the trapped fractional mode density is quantized in units of $\frac{1}{4}$, a quantization unique to HOTIs \cite{FCA}, and scales proportionally to the Frank angle of the disclinations \cite{CornerCharge,tianheDisclin,LiuShift2019}, as predicted by Eqn.~\eqref{disc_index} (the term proportional to polarization always contributes integer mode density for the type of disclinations we study, see \ref{disclination}).
Although fractional mode density is typically associated with topological bound states \cite{FCA}, we observe that no topological states are trapped by the disclinations when the fractional mode density lies within the defective central/core unit cells.
For each case, we then locally deform the area surrounding the disclination core while keeping the overall rotation symmetry, such that the defective unit cell is decoupled from the rest of the lattice and the trapped fractional mode density symmetrically splits and moves away from the core.
Remarkably, these local deformations reveal an odd number of in-gap topological states trapped by each disclination defect.
We then further deform both metamaterial samples and break rotation symmetry such that pairs of these in-gap states are strongly coupled to each other and become gapped, merging into the bulk bands. This final deformation leaves only a single defect-induced bound state remaining in the bandgap.
Our results conclusively show that disclination defects in TCIs can trap higher-order fractional charges as well as topological bound states. 
Furthermore, our experiments demonstrate the robustness of \emph{fractional charge} as a key topological indicator, in contrast to the existence of in-gap topological bound states, which may not always manifest at disclination defects. 

\vspace{12 pt}

The $C_4$-symmetric HOTI metamaterials in this study are based on a simple tight-binding model where each unit cell contains four atoms.
Each atom is strongly bonded to its nearest neighbors in adjacent unit cells, as shown in Fig.~\ref{fig1}a, and is weakly bonded to its nearest neighbors within the same unit cell.
This HOTI has four bulk bands separated by two bandgaps (the central two bands are degenerate), gapped edge bands, and topological corner states \cite{Mittal2019, BadHOTI, FCA}.
Most importantly for our study, at $\frac{1}{4}$ filling (i.e. only the lowest-energy bulk band is filled) the HOTI exhibits a fractional charge of $\frac{1}{2}$ at edges and $\frac{1}{4}$ at corners \cite{CornerCharge, FCA}.

The creation of disclination defects through a ``cutting and gluing" procedure is illustrated in Fig.~\ref{fig1} for the HOTI described above, for disclinations with both negative and positive Frank angles.
First, the lattice is cut into four identical sectors, which is always possible due to the $C_4$ symmetry.
Since the bulk of these sectors is identical to the original insulator, each sector also has a $\frac{1}{2}$ edge fractional charge and $\frac{1}{4}$ corner fractional charge through the bulk-boundary correspondence.
A disclination defect with a negative Frank angle can be formed by removing one sector and then gluing the three remaining pieces back together, forming a nominally triangular shape (Fig.~\ref{fig1}b).
The Frank angle $\Omega$ can be calculated by drawing a closed loop around the disclination, as illustrated in the schematic of Fig.~\ref{fig1}b (red arrows) and described in \ref{disclination}. 
Here, a closed loop around the defect needs only three $90^{\circ}$ turns, one less than a closed loop in a defect-free lattice, giving a negative Frank angle $\Omega = -90^{\circ}$.
A disclination with a positive Frank angle can be formed by adding one sector, forming a pentagonal shape (Fig.~\ref{fig1}c).
In this case, a closed loop around the defect requires five $90^{\circ}$ turns, giving a positive Frank angle $\Omega = +90^{\circ}$. 
We note that although the HOTI has non-zero electric polarization at $\frac{1}{4}$ filling, the contribution to fractional charge from the second term in Eqn.~\eqref{disc_index} is zero for both disclinations; both  are characterized by the same Burgers vector (see \ref{disclination}).
We therefore expect the trapped charge at the disclinations to be directly proportional to their Frank angle.
When all four sectors are combined the fractional boundary charges sum to an integer everywhere within the bulk, leaving a uniform integer charge density. 
However, at disclination defects, an odd number of sectors are combined, leading to leftover fractional charge at the disclination core.
The necessity for this leftover charge can also be seen from the exterior of the system, which has only three (or five) exterior corners due to the defect.
In general, the total charge of any insulator must take an integer value since the number of filled states is always an integer.
Here, because each corner carries a $\frac{1}{4}$ fractional charge, the odd number of corners contribute an overall fractional charge of $\pm \frac{1}{4}$ to the insulator, which can only be compensated by a $\frac{1}{4}$ fractional charge in the bulk.
We note this $\frac{1}{4}$ fractional charge trapped by the disclinations is a feature unique to \emph{higher-order} topological insulators \cite{FCA}, since the first-order features in $C_4$ symmetric TCIs, namely the polarization and related edge fractional charge, are always quantized in units of $\frac{1}{2}$.

\vspace{12 pt}

\begin{figure}
	\begin{adjustwidth}{-1in}{-1in}
    \centering
    \includegraphics[width = \linewidth]{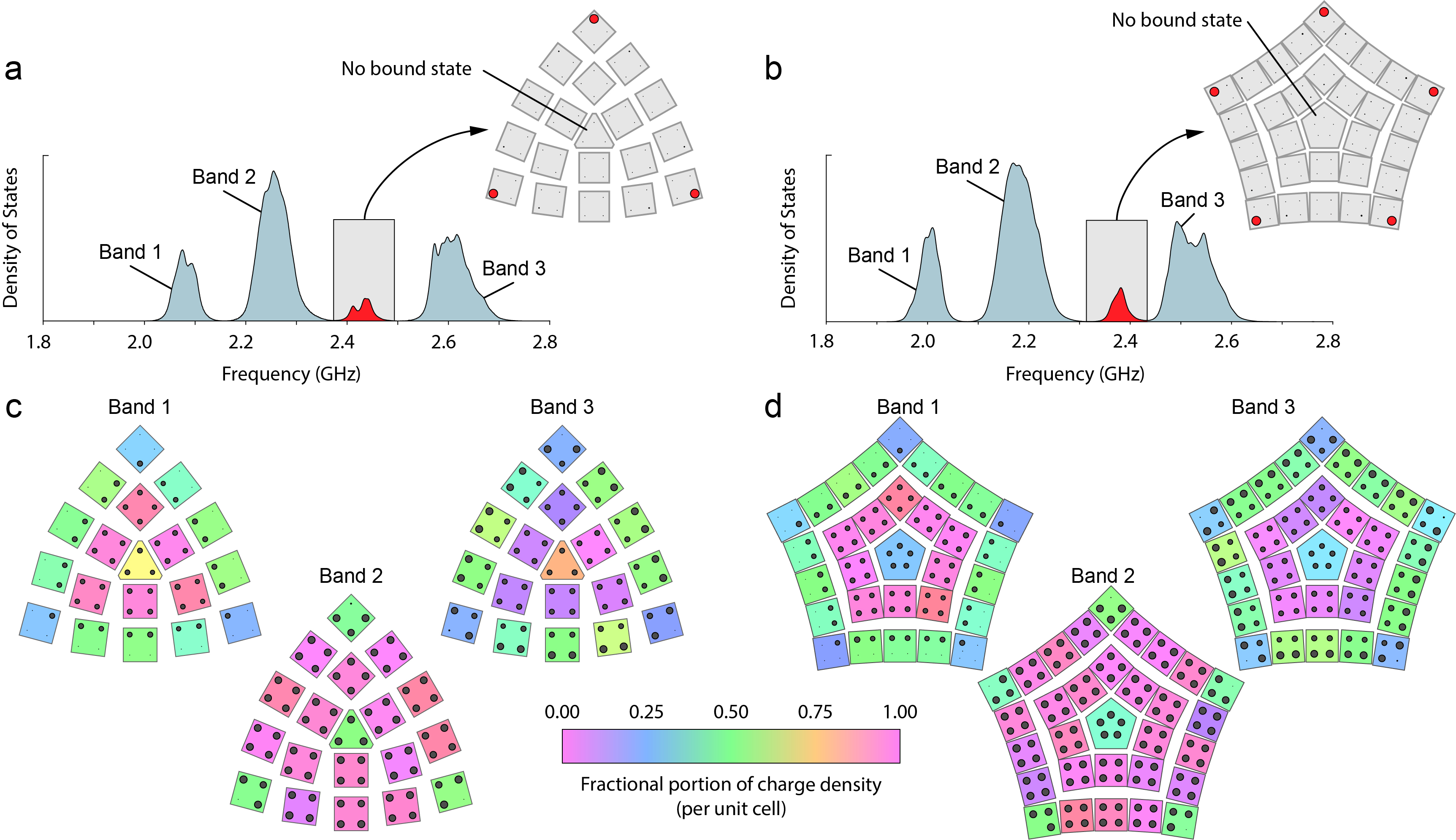}
    \caption{Measurement of trapped fractional mode density. (a) Measured density of states (DOS) for the $C_3$-symmetric board in Fig. \ref{fig1}c. The highlighted in-gap states are localized to the three corners, with no in-gap state at the disclination core. (b) Same as (a) but for the $C_5$-symmetric board in Fig. \ref{fig1}d, which has five in-gap corner states. (c) Measured mode density per unit cell in each of the three bands of the $C_3$-symmetric board. The color of each unit cell indicates the fractional portion of mode density in that cell, and the area of each dot is proportional to the mode density of the corresponding resonator. Here, the central unit cell has approximately $\frac{3}{4}$ fractional mode density in bands (1) and (3). (d) Same as (c) but for the $C_5$-symmetric board. Here, the central unit cell has approximately $\frac{1}{4}$ fractional mode density in bands (1) and (3). Detailed versions of (c) and (d) showing the total mode density per unit cell are available in \ref{bigfigs}.}
    \label{fig2}
    \end{adjustwidth}
\end{figure}

We physically realize the HOTIs with disclination defects in metamaterials consisting of coupled microstrip resonators, each with a fundamental resonance frequency $f_0 \approx 2.6$~GHz and a quality factor $Q \approx 160$.
Each resonator corresponds to an atom, or a single degree of freedom, and the resonators are coupled by discrete capacitors that correspond to the bonds between atoms.
The fabricated microwave-frequency metamaterials with negative and positive disclination angles are respectively shown in Fig.~\ref{fig1}d,e.
Note that the physical coupling within the bulk forms a symmetric square in both metamaterials, as shown in the insets of Fig.~\ref{fig1}d,e and schematically in Fig.~\ref{fig1}b,c, except at the center of each board.
In order to preserve this square coupling region, which helps to ensure equal coupling rates between resonators, the shapes of the individual resonators become distorted toward the edges of the board. 
Nevertheless, all resonators are designed to have the same electric length and as such have the same fundamental resonance frequency of $2.6$~GHz. 
The metamaterials are also designed to have an overall $C_3$ or $C_5$ symmetry, as this global rotation symmetry fixes the total fractional mode density in each symmetric sector.
To characterize the fabricated metamaterials, we first measure the spectral DOS of each site (each resonator) individually using a network analyzer reflection measurement (see Methods).
The total DOS spectrum, calculated by summing over all sites, is shown in Fig.~\ref{fig2}a,b for the $C_3$- and $C_5$-symmetric metamaterials respectively.
Since both are based on the same tight-binding Hamiltonian and only differ in the Frank angle of their disclination defect, we expect, and indeed find, that the spectral DOS are nearly identical.
For both insulators we measure three large, well-defined bands and a small number of states within the largest bandgap.
The in-gap states are strongly localized to the materials' corners (Fig.~\ref{fig2}a,b) and lie within the same unit cell as a $\frac{1}{4}$ fractional mode density (Fig.~\ref{fig2}c,d), indicating that these are the topological corner states of the HOTI \cite{FCA}.
Later, we will show that the fractional charge at the disclination core can also be associated with topological bound states, although these states are missing when the fractional mode density is within the defective unit cell at the disclination core, as in Fig.~\ref{fig2}a,b.
We note that a small on-site potential is added to the corner sites in order to shift these corner states into the bulk bandgap, otherwise these states would be obscured by the bulk bands.
We now shift focus to the spatial distribution of mode density within the three large bands, concentrating on the fractional part.
The spatially resolved mode densities for each band of the $C_3$- and $C_5$-symmetric metamaterials, which have an identical $C_4$-symmetric unit cell structure, are shown in Fig.~\ref{fig2}c,d respectively.
Since there are no energetic filling rules in these systems, we can consider the mode density distribution of each band separately.
In the lowest and highest frequency bands, the edge unit cells have a fractional mode density of approximately $\frac{1}{2}$, and each corner unit cell carries about $\frac{1}{4}$ fractional mode density.
The central band of this HOTI has identical density features as the other bands but is doubly degenerate, which doubles the mode density when compared to the singly degenerate bands \cite{FCA}.
As such, in the central band only the corner unit cells have a fractional mode density (a value of $\frac{1}{2}$).
In the bulk unit cells, the mode density takes an integer value for all bands, except in the central unit cell where the disclination occurs.
For the negative disclination ($\Omega = -90^{\circ}$), this defective central unit cell has a fractional mode density of $\frac{3}{4}$ in the singly degenerate bands, as predicted by the gluing picture in Fig.~\ref{fig1}.
In contrast, the positive disclination ($\Omega = +90^{\circ}$) generates a fractional mode density of $\frac{1}{4}$ in the defective unit cell for these bands, which is also expected.
We note that these trapped charges are also doubled in the twice-degenerate central band.

\vspace{12pt}

Having established the existence of trapped fractional mode density at the disclination cores, we will now demonstrate its relationship to trapped topological bound states.
Boundary-induced fractional mode density has been shown to be associated with topological boundary states \cite{FCA}, as exemplified by the corner unit cells of both insulators in Fig.~\ref{fig2}.
However, for the reason detailed below, we expect that the association between fractional mode density and topological states only holds when all unit cells in the system contain the same number of sites/modes.
The fractional mode density is a collective property of the bulk and cannot be changed by deformations, even relatively violent ones that locally add or remove sites, as we show in \ref{rmvsite}, but such deformations \emph{can} remove topological bound states.
This property is clearly evident in the zero-correlation-length limit, where the weak coupling rate within unit cells goes to 0.
In this limit topological bound states will lie entirely on sites that are decoupled from the rest of the system. As such, removing these sites will also remove the bound states.
Since the zero-correlation-length limit can always be deformed into an arbitrary \emph{finite} correlation length without closing the bandgap or changing the number of bound states, this property is also true in general (see discussion in \ref{rmvsite}).
\begin{figure}
	\begin{adjustwidth}{-1in}{-1in}
    \centering
    \includegraphics[width = 0.7\linewidth]{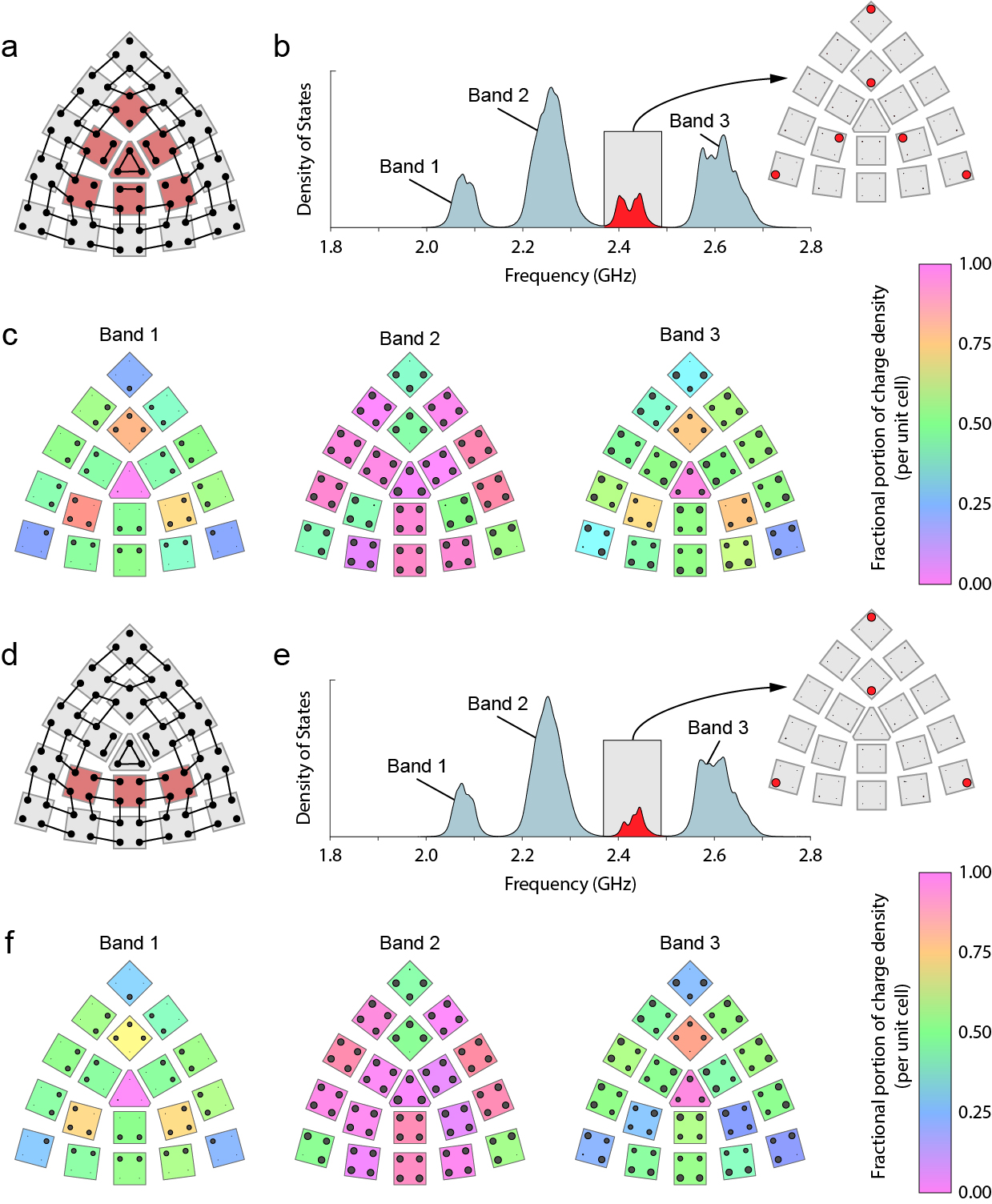}
    \caption{Local deformation reveals trapped in-gap state ($C_3$). (a) Deformed $C_3$-symmetric lattice with trivial central unit cell (only highlighted unit cells are deformed). (b) Measured DOS for the deformed lattice in (a). Highlighted in-gap states are localized to the original exterior corners, plus three interior corners. (c) Measured mode density for each of the three bands shown in (b). The central unit cell now has an integer mode density, and surrounding unit cells host fractional mode density. (d-f) Same as (a-c) but the lattice is further deformed, breaking $C_3$ rotation symmetry, such that two of the interior corner states are gapped out. Highlighted unit cells in (d) are deformed from (a). The fractional mode density in bands (2) and (3) changes in these unit cells due to the deformation. Detailed versions of (c) and (f) showing the total mode density per unit cell are available in \ref{bigfigs}.}
    \label{fig3}
    \end{adjustwidth}
\end{figure}
\begin{figure}
	\begin{adjustwidth}{-1in}{-1in}
    \centering
    \includegraphics[width = 0.7\linewidth]{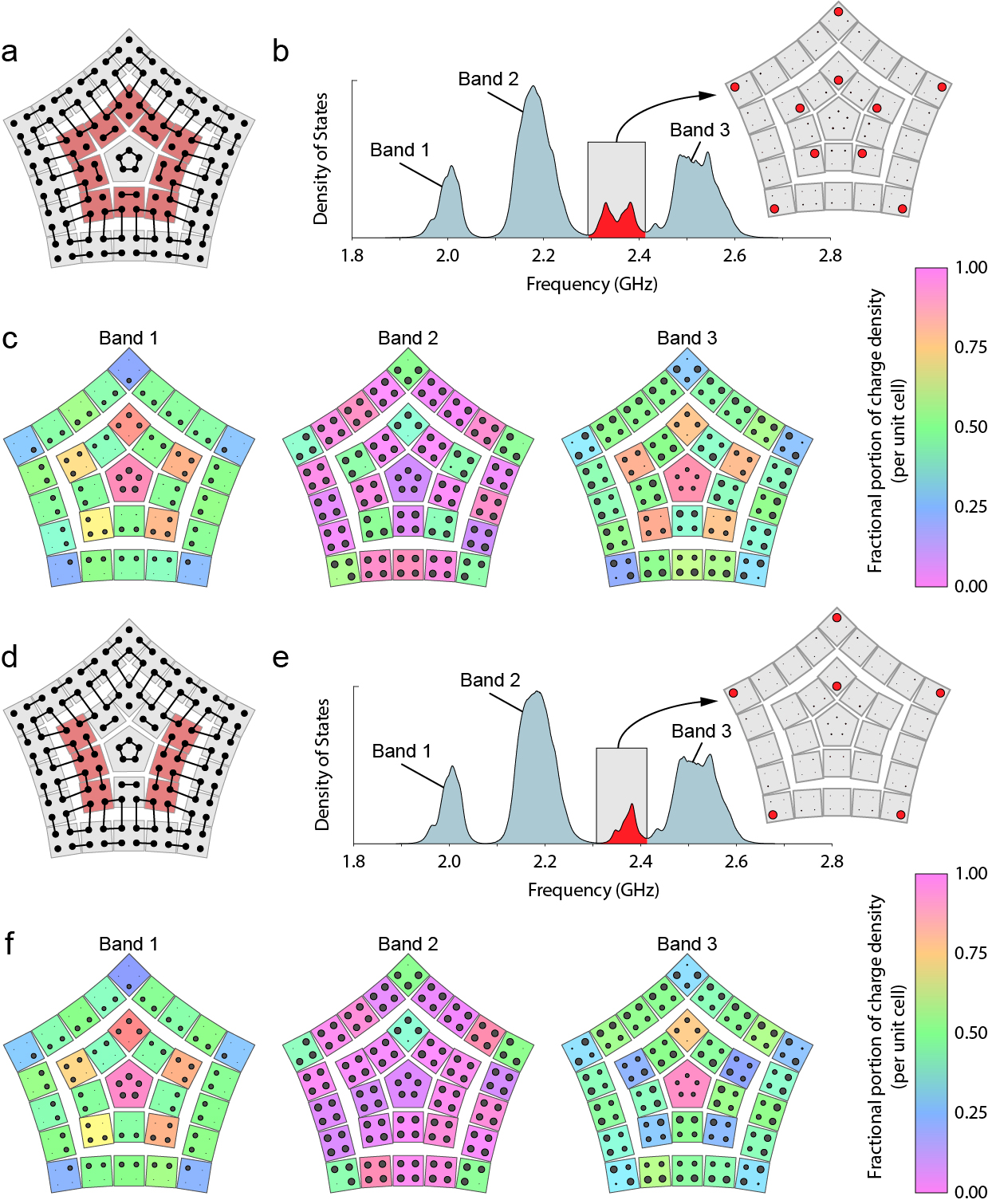}
    \caption{Local deformation reveals trapped in-gap state ($C_5$). (a-f) Same as Fig.~\ref{fig3} but for the $C_5$-symmetric insulator. Detailed versions of (c) and (f) showing the total mode density per unit cell are available in \ref{bigfigs}.}
    \label{fig4}
    \end{adjustwidth}
\end{figure}
In our fabricated metamaterials the unit cell contains four sites, but the defective central unit cell has three sites in the $C_3$-symmetric insulator and five in the $C_5$-symmetric insulator.
Hence, we do not expect a conventional association between the trapped fractional mode density and topological bound states.
This is supported by the measurements in Fig.~\ref{fig2}, which show no evidence of bound states at the disclinations --- the spectrum of each site in the bulk is identically gapped, leaving no room for a localized bound state.
However, although neither of these central unit cells hosts topological bound states, we will now show that such states can be generated near the disclination core by deformations that shift the fractional charge from the defective unit cell into the surrounding intact unit cells.
We note that these newly generated bound states are not added to the system through such deformations, but naturally appear due to the topological nature of the disclination defects when the defective unit cell at the core is removed.
We first deform the central unit cell in both metamaterials such that the sites within it are strongly coupled to each other and are weakly coupled to sites in neighboring unit cells. This is equivalent to putting \emph{only the central} unit cell in a topologically trivial phase.
We also deform the unit cells immediately adjacent to the disclination such that they form an interior boundary, similar to to the exterior boundary, around the central unit cell.
These deformations do not break the global rotation symmetry, as illustrated in the schematics of the deformed $C_3$- and $C_5$-symmetric insulators respectively shown in Figs.~\ref{fig3}a and \ref{fig4}a.
After the deformations are complete, we again measure the spectral DOS of each site to find the spatial mode density, which is shown in Figs.~\ref{fig3}c and \ref{fig4}c for the $C_3$- and $C_5$-symmetric materials respectively.
As the central unit cell of both metamaterials is now topologically trivial, the mode density in this unit cell takes an integer value in all three bands.
We find that the fractional mode density that was previously trapped at the disclination has symmetrically split and moved to the interior boundary, with $\frac{3}{4}$ charges on each interior corner, and $\frac{1}{2}$ along the interior edge in the singly degenerate bands.
In the $C_3$-symmetric insulator, the original $\frac{3}{4}$ mode density splits, creating three unit cells with $\frac{3}{4}$ mode density and three with $\frac{1}{2}$ mode density (total $3 \frac{3}{4}$).
Likewise, the $C_5$-symmetric insulator, which originally had a $\frac{1}{4}$ mode density in the center unit cell, contains five unit cells with $\frac{3}{4}$ mode density and five with $\frac{1}{2}$ mode density after the deformation (total $6 \frac{1}{4}$).
We note that the total \emph{fractional} part of the mode density (i.e., the mode density modulo 1) in each sector  remains constant for each of the bands, even after these deformations, since we do not break the global rotation symmetry and the bulk bandgap does not close away from the core.
In addition to the splitting of fractional mode density, this deformation also results in three (or five) additional topological bound states within the bandgap.
These states are located at the interior corners of the system around the trivial central unit cell, as shown in Figs.~\ref{fig3}b and \ref{fig4}b respectively.
We note that there are an odd number of bound states that emerge in both deformed systems, which is required by the global $C_3$ or $C_5$ rotation symmetry since the bound states must either lie at the rotation center or be identically distributed in each of the three (or five) symmetric sectors.
These states are not protected against deformations that break these symmetries, however, even if the rotation symmetry is broken, the topologically non-trivial nature of these disclination defects ensures that one bound state always survives (except in the cases where the unit cell is defective and broken, as discussed above).
We demonstrate this property by breaking the rotation symmetry through further local deformation of both metamaterial samples, as illustrated schematically in Figs.~\ref{fig3}d and \ref{fig4}d. Note that only the highlighted unit cells are deformed.
This deformation strongly couples \emph{all but one} of the disclination-induced bound states in pairs.
The coupled states become gapped and no longer lie within the bandgap (Figs.~\ref{fig3}e and \ref{fig4}e), instead splitting such that they enter both the central and upper bulk bands.
This is accompanied by an increase of $\frac{1}{2}$ fractional mode density in each of these bands within the highlighted interior-corner unit cells (Figs.~\ref{fig3}f and \ref{fig4}f).
We note that this increase is only possible because the rotation symmetry is broken by the deformation, and the fractional mode density in each sector is no longer symmetry-protected.
Following these deformations, a single topological bound state remains within the bulk of each deformed insulator, as shown in Figs.~\ref{fig3}e and \ref{fig4}e.
Since the total fractional mode density localized on each disclination does not change during any of our deformations, we can associate these singular bound states with the disclination-induced fractional mode density. 

\vspace{12pt}

Topological bound states trapped by disclinations could prove useful for a variety of engineering applications since these defects can lie deep within the bulk away from the material boundaries.
For example, topological pumps have been shown to be capable of robustly transferring vibrational energy between topological bound states \cite{pumpingpaper}.
Similar pumping processes could be used to transfer energy into bound states trapped at disclinations deep within the bulk of a material, where the energy could be safely stored without radiative decay due to the surrounding insulator.
Topological edge states have also been shown to exhibit desirable properties for laser applications, including high-power single mode lasing, robustness against fabrication defects, and high slope efficiency \cite{Bandreseaar4005, Hararieaar4003, StJeanLaser}.
Defect-bound topological states could be used to create similar topological lasers (especially in 3D HOTIs), which could prove less sensitive to external noise and radiative loss since they are not fixed to a surface.
Furthermore, experiments on certain types of non-linear topological insulators have shown that topological bound states can be used to drive topological transitions \cite{HadadNL, HadadNLExp}, including in HOTIs \cite{FleuryNL}.
Our results provide an important first step to these applications, demonstrating that topological features trapped at disclinations can be identified through measurement of charge density and, if necessary, revealed through local deformations.

\section*{Acknowledgements}

The authors would like to thank Prof. Jennifer T. Bernhard for access to the resources at the UIUC Electromagnetics Laboratory. This project was supported by the US National Science Foundation (NSF) Emerging Frontiers in Research and Innovation (EFRI) grant EFMA-1627184. C.W.P. additionally acknowledges support from the NSF Graduate Research Fellowship. G.B. additionally acknowledges support from the US Office of Naval Research (ONR) Director for Research Early Career Grant N00014-17-1-2209. T.L. and T.L.H. additionally thank the U.S. National Science Foundation under grant DMR-1351895. 

\vspace{12pt}

\section*{Author contributions}

C.W.P. designed and fabricated the microwave circuits, performed the microwave simulations and experimental measurements, and produced the experimental figures. W.J. assisted the microwave circuit design and experimental measurements. T.L. guided the topological insulator design and performed the theoretical calculations. T.L.H. and G.B. supervised all aspects of the project. All authors jointly wrote the paper.

\section*{Methods} 
\label{methods}
\section*{Measuring density of states in a microwave metamaterial} 
We experimentally find the local DOS of the microwave metamaterials by first measuring the reflection spectrum $S_{11}(f)$ at each resonator, where $f$ is the frequency.
The reflection measurements are taken using a microwave network analyzer (Keysight E5063A). 
The reflection probe is composed of a 50 $\Omega$ coaxial cable terminated in a $0.1$ pF capacitor, which is contacted to each resonator at an anti-node. 
Due to the low probe capacitance, the measured linewidths are dominated by intrinsic losses in each resonator.
The background reflection contributed by the probe is evaluated away from any modes and is removed.
This measurement process is similar to that used in Ref. \cite{Peterson2018}.
The absorptance $A(f)$, which is defined as the ratio of absorbed power to incident power, can be calculated from the reflection as $A(f) = 1 - \left| S_{11}(f) \right| ^2$.
To obtain the density of states $D(f)$ for each resonator, we divide the measured absorptance by the frequency squared, $D(f) = A(f) / f^2$, which accounts for increased coupling to the capacitive probe at higher frequencies.
Finally, we normalize $D(f)$ such that $$\int_{\text{all bands}} D_{\textbf r}(f) = 1,$$ where the integration is over the whole band structure and $D_{\textbf r}(f)$ is the local density of states for one resonator, indexed by $\textbf r$. 
Since each resonator supports a single mode within the measured frequency range, such that for an $N$ resonator system there are $N$ modes total, this normalization maps each mode to a mode density of $1$.

\section*{Design of microwave-frequency metamaterials} 
The metamaterials are fabricated on $0.787$ mm thick Rogers RT/duroid 5880 substrate, with $35~\mu$m thick copper on each side.
The resonators that make up each metamaterial are half-wavelength microstrip transmission lines, with a characteristic impedance of each section $Z_0 \approx 110~\Omega$. The resonator layout is such that the coupling between resonators forms a square (see Fig.~\ref{fig1}), requiring that the resonator shape changes slightly from the center of each board to the outside. To keep the electrical length of each resonator approximately equal, the resonators have a curved section in their center, where the curvature decreases away from the center of the $C_3$-symmetric board, and increases away from the center of the $C_5$-symmetric board.
While there are losses in both the dielectric substrate and copper conductor, as well as comparatively insignificant radiative losses, these are small (the typical resonator linewidth is $\approx 16$~MHz, for a typical quality factor of $\approx 160$) and do not affect the underlying topology.
The coupling between resonators is implemented using two discrete capacitors in series, such that the strong coupling capacitance is $0.3$~pF (two $0.6$~pF capacitors), and the weak coupling capacitance is $0.05$~pF (two $0.1$~pF capacitors).

\newpage

\newcommand{\beginsupplement}{%
        \setcounter{table}{0}
        \renewcommand{\thetable}{S\arabic{table}}%
        \setcounter{figure}{0}
        \renewcommand{\thefigure}{S\arabic{figure}}%
        \setcounter{equation}{0}
        \renewcommand{\theequation}{S\arabic{equation}}%\
        \setcounter{section}{0}
        \renewcommand{\thesection}{S\arabic{section}}%
}

\beginsupplement

\begin{center}
\Large{\textbf{Supplementary Information: \\ Observation of trapped fractional charge and topological states at disclination defects in higher-order topological insulators}} \\
\vspace{12pt}
\vspace{12pt}
\large{{Christopher W Peterson}$^1$,
{Tianhe Li}$^2$,
{Wentao Jiang}$^1$,
{Taylor L. Hughes}$^2$,
and {Gaurav Bahl}}$^3$ \\
\vspace{12pt}
{\footnotesize{$^1$Department of Electrical and Computer Engineering, University of Illinois at Urbana-Champaign, Urbana, IL, USA}} \\
{\footnotesize{$^2$Department of Physics and Institute for Condensed Matter Theory, University of Illinois at Urbana-Champaign, Urbana, IL, USA}} \\
{\footnotesize{$^3$Department of Mechanical Science and Engineering, University of Illinois at Urbana-Champaign, Urbana, IL, USA}}\\
\end{center}

\section{Index of fractional disclination charge}
\label{disclination}
\begin{figure}[h!]
\centering
\includegraphics[width=\textwidth]{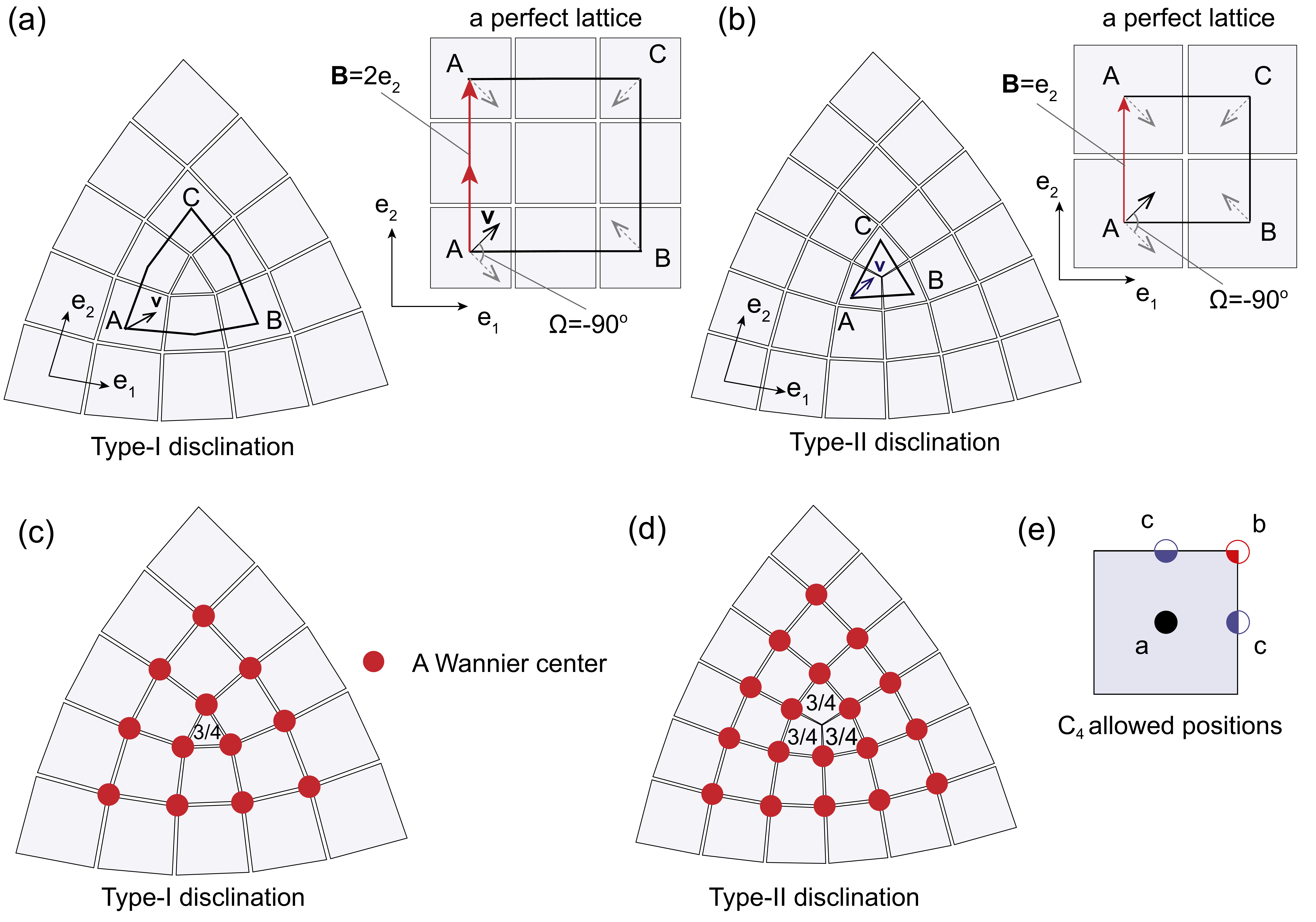}
\caption{(a, b) Demonstration of calculating the Burgers vector and Frank angle for a $C_4$ symmetric square lattice having the (a) Type-I disclination, (b) Type-II disclination. The black arrows indicate the vector being parallel transport enclosing the disclination core and the dahsed gray arrows indicate that vector in the process of the parallel transport. (c, d) The Wannier center configuration and fractional charge in defective lattices having the (c) Type-I disclination, (d) Type-I disclination. (e) Allowed positions of Wannier centers by the $C_4$ symmetry in one unit cell.}
\label{fig:disclination_lattice}
\end{figure}

In this section we describe in detail the quantities in Eqn. \eqref{disc_index} in the Main Text. We also discuss why the fractional charges trapped by the disclinations studied in the Main Text are proportional to their Frank angle.
A disclination defect is characterized by the amount of net translation (denoted by the Burgers vector $\mathbf{B}$) and net rotation (denoted by the Frank angle $\Omega$ ) accumulated under parallel transport of a vector along a loop enclosing the disclination core.
To illustrate these two quantities, we consider a $C_4$ symmetric square lattice having a disclination defect at the center.
As shown in Fig.~\ref{fig:disclination_lattice}(a), the lattice vectors are denoted by ${\bf e}_1, ~{\bf e}_2$.
The vector ${\bf v}$ is parallel transported along the loop $ABCA$ that encloses the disclination core, meaning that the vector is translated around the loop by an integer number of lattice vectors with a $90^\circ$ rotation at each corner.
In comparison to the same transport conducted in a perfect lattice without defects, which is shown in the inset of Fig.~\ref{fig:disclination_lattice}(a), we find that after enclosing the disclination core the final vector has gained an extra rotation of $-90^{\circ}$,  and an extra translation of $2{\bf e}_2$, corresponding to $\Omega =-90^{\circ}, \, {\bf B}=2{\bf e}_2$. 
In two dimensional $C_4$ symmetric lattices, the parity of the sum of the Burgers vector components classifies disclinations having the same Frank angles \cite{TeoDisclin}, such that for a fixed Frank angle there are only two types of topologically distinct disclinations.
In type-I disclinations the sum of the Burgers vector components is even, as shown in Fig.~\ref{fig:disclination_lattice}(a).
A type-II disclination, where the extra translation accumulated after looping around the disclination core is $1 {\bf e}_2$, is shown in Fig.~\ref{fig:disclination_lattice}(b).
The two different types of disclinations trap different fractional charges, even if they have the same Frank angle. As shown recently \cite{tianheDisclin}, one can find the fractional disclination charge by inspecting the real-space localized Wannier representation of an insulator. 
The representation corresponding to the $C_4$-symmetric HOTI studied in the Main Text has a single Wannier center at the corner of a unit cell (Wyckoff position $b$), which generates a bulk polarization ${\bf P} =\frac{1}{2}({\bf e}_1 +{\bf e}_2)$ (we set the charge $e=1$). 
Due to the $C_4$ symmetry, the single Wannier center at the unit cell corner contributes an equal quantized fractional charge of $\frac{1}{4}$ to each of the adjacent four unit cells. 
In Fig.~\ref{fig:disclination_lattice}(c,d), we show how the bulk Wannier centers are arranged around both type-I and type-II disclinations. 
For the lattice having a type-I disclination, the unit cell at the core receives contributions from three Wannier centers, leading to a fractional charge of $\frac{3}{4}$.
In contrast, for the lattice having a type-II disclination, each of the three unit cells around the disclination core manifests a fractional charge of $\frac{3}{4}$, leading to a total fractional disclination charge of $\frac{9}{4} \mod{1}=\frac{1}{4}$. 
In addition to the corner of the unit cell (Wyckoff position $b$), Wannier centers in a $C_4$ symmetric insulator are allowed at all the positions shown in Fig.~\ref{fig:disclination_lattice}(e).
For these more general cases, the index of fractional disclination charge in $C_4$ symmetric insulators is \cite{tianheDisclin}
\begin{align}
Q_{dis}= \frac{\Omega}{2\pi}(n_b +2n_c) + \sum_{i,j=1,2}\epsilon_{ij}B_iP_j \mod 1
\label{eq:c4_index}
\end{align}
where $n_b$ and $n_c$ are respectively the number of Wannier centers located at the Wyckoff positions $c$ and $b$, which are determined by the band topology of the insulator; $B_i$ and $P_i$ are respectively the component of the Burgers vector and the bulk polarization; and $\epsilon_{ii}=0, \epsilon_{ij}=-\epsilon_{ji}=1$. 
Notice that for type-I disclinations in $C_4$ symmetric insulators, since the sum of components of the Burgers vector is always even and $P_1 = P_2$, the second term in Eq.~\eqref{eq:c4_index} always gives an integer number.
Therefore, the fractional disclination charge index for the type-I disclination is,
\begin{align}
Q_{dis, type-I}=\frac{\Omega}{2\pi}(n_b +2n_c) \mod 1.
\end{align}
Therefore, the fractional charge trapped at the disclination core for a given band structure is only proportional to the Frank angle $\Omega$, which is consistent with our experimental measurement.

\section{Effects of removing sites from the lattice}
\label{rmvsite}
\begin{figure}[h!]
\centering
\includegraphics[width=\textwidth]{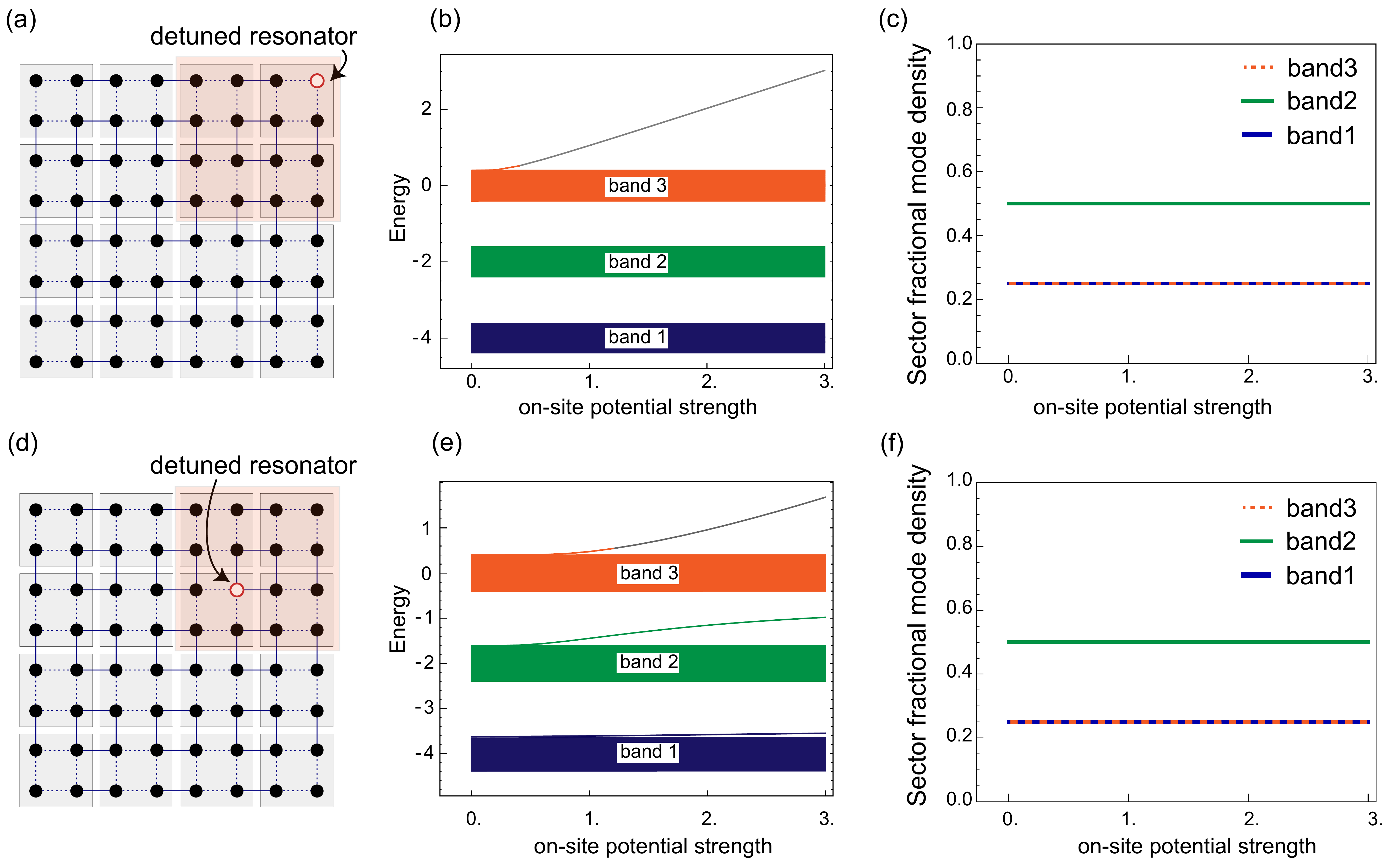}
\caption{(a, d) Lattice configuration of the HOTI used in simulations. We consider a lattice having $ 10\times 10$ unit cells, intra-cell coupling strength of $0.25$ (dashed lines) and inter-cell coupling strength of $1$ (solid lines). The resonator being removed is denoted by a red circle. (b, e) The spectrum of the lattice in (a, d) as a function of the on-site potential applied to the highlighted resonators. (c, e) The fractional part of the mode density integrated over the shaded sector in (a, f) for each band.}
\label{fig:supp_defectiveUC}
\end{figure}

In this section, we show through simulations that removing a site from the lattice will not change the total fractional mode density, but can remove topological bound states. 
Let us consider the same $C_4$-symmetric HOTI as studied in the Main Text, the lattice configuration of which is shown (with no defects) in Fig.~\ref{fig:supp_defectiveUC}(a,d).
To mimic our experiments, we intentionally do not compensate the intrinsic capacitive loading effect \cite{FCA}, which shifts the energy of the bulk bands down in relation to the energy of the topological corner states (Fig.~\ref{fig:supp_defectiveUC}(b,e).
The resulting system has three bands as shown in Fig.~\ref{fig:supp_defectiveUC}(b,e), and can be divided into four identical sectors due to the $C_4$ symmetry.
The singly-degenerate bands, band $1$ and band $3$, manifest a fractional mode density of $\frac{1}{4}$ in each sector, while the doubly-degenerate band $2$ manifests that of $\frac{1}{2}$ per sector.
To simulate removing a site adiabatically, we apply an increasing on-site potential to that site.
We first consider detuning the resonator at the upper right corner, which hosts a localized topological corner mode. 
As shown in Fig.~\ref{fig:supp_defectiveUC}(b), the corner mode is lifted and separated from the band structure as the on-site potential increases and the corner site is effectively removed from the lattice.
However, it is evident from Fig.~\ref{fig:supp_defectiveUC}(c) that the fractional part of mode density in each band, integrated over the whole top-right sector, remains constant through the whole process.
This occurs because an integer number of modes are removed from that sector, such that the mode density can only change by an integer number.
Furthermore, the fractional part of the total mode density is a collective behavior of all modes the in the band, and with only nearest-neighbor couplings a local deformation does not affect the mode density in the areas far away from that deformation.
We also consider removing a bulk resonator in the upper right sector, the resulting spectrum and sector fractional mode density are shown in Fig.~\ref{fig:supp_defectiveUC}(e,f).
As before, one mode, localized where the on-site potential is applied, is dragged out of the band structure while the integrated fractional mode density in the top-right sector remains the same.
Additionally, note that this deformation introduces one mode into the bandgap, thereby creating an in-gap bound state despite not changing the fractional mode density.

\section{Detailed experimental figures}
\label{bigfigs}
In this section, we show more detailed versions of the experimental figures in the main manuscript, showing explicitly the total measured mode density for each unit cell.
\begin{figure}[h!]
\begin{adjustwidth}{-1in}{-1in}
\centering
\includegraphics[width=\linewidth]{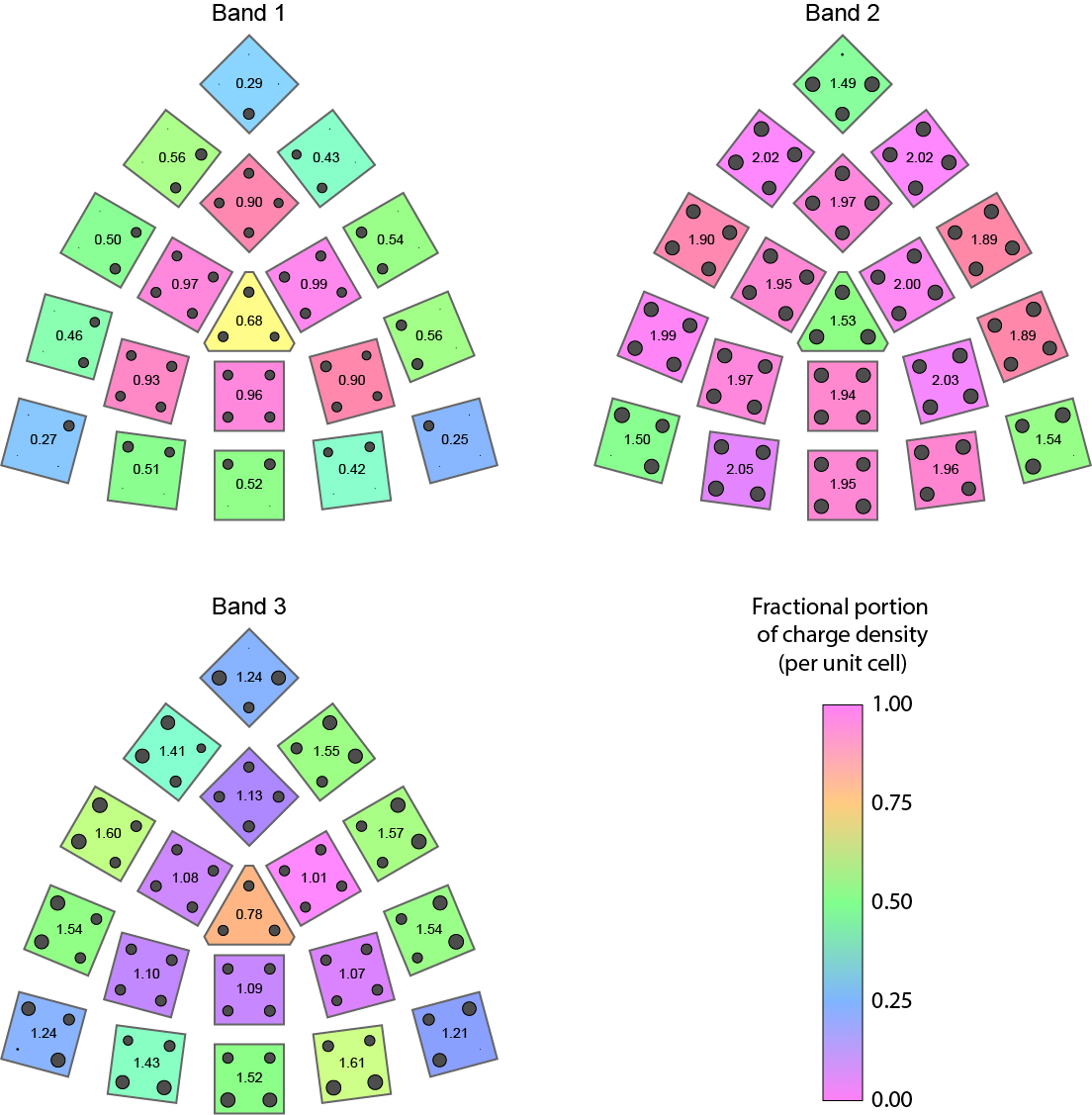}
\caption{Measured mode density for the $C_3$-symmetric insulator, as shown in Fig. \ref{fig2}(c). The total mode density for each band is listed numerically in each unit cell, and the area of each dot is proportional to the mode density of the corresponding resonator. The color of each unit cell indicates the fractional portion of mode density.}
\label{figs3}
\end{adjustwidth}
\end{figure}

\begin{figure}[h!]
\begin{adjustwidth}{-1in}{-1in}
\centering
\includegraphics[width=\linewidth]{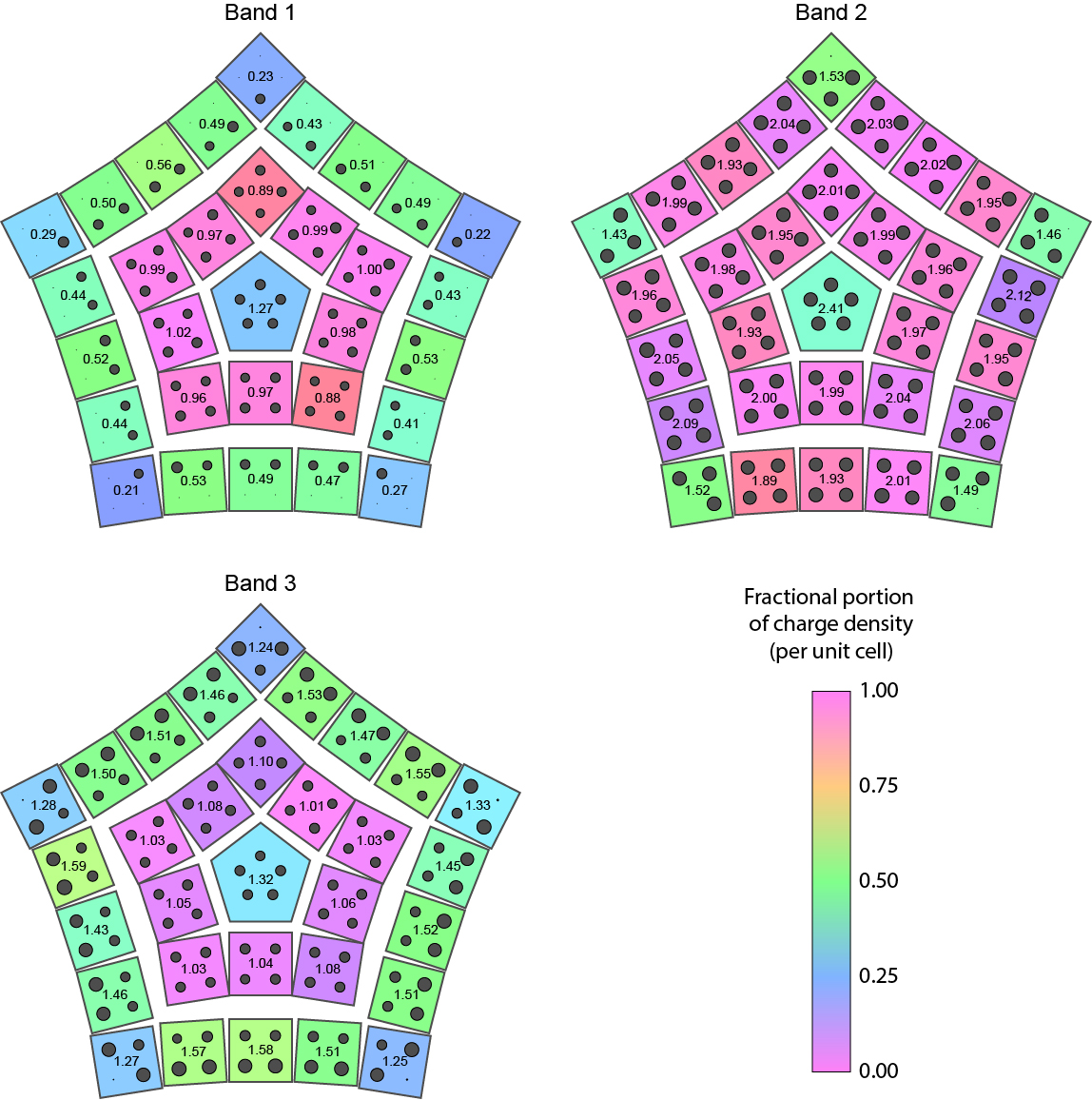}
\caption{Measured mode density for the $C_5$-symmetric insulator, as shown in Fig. \ref{fig2}(d). The total mode density for each band is listed numerically in each unit cell, and the area of each dot is proportional to the mode density of the corresponding resonator. The color of each unit cell indicates the fractional portion of mode density.}
\label{figs4}
\end{adjustwidth}
\end{figure}

\begin{figure}[h!]
\begin{adjustwidth}{-1in}{-1in}
\centering
\includegraphics[width=\linewidth]{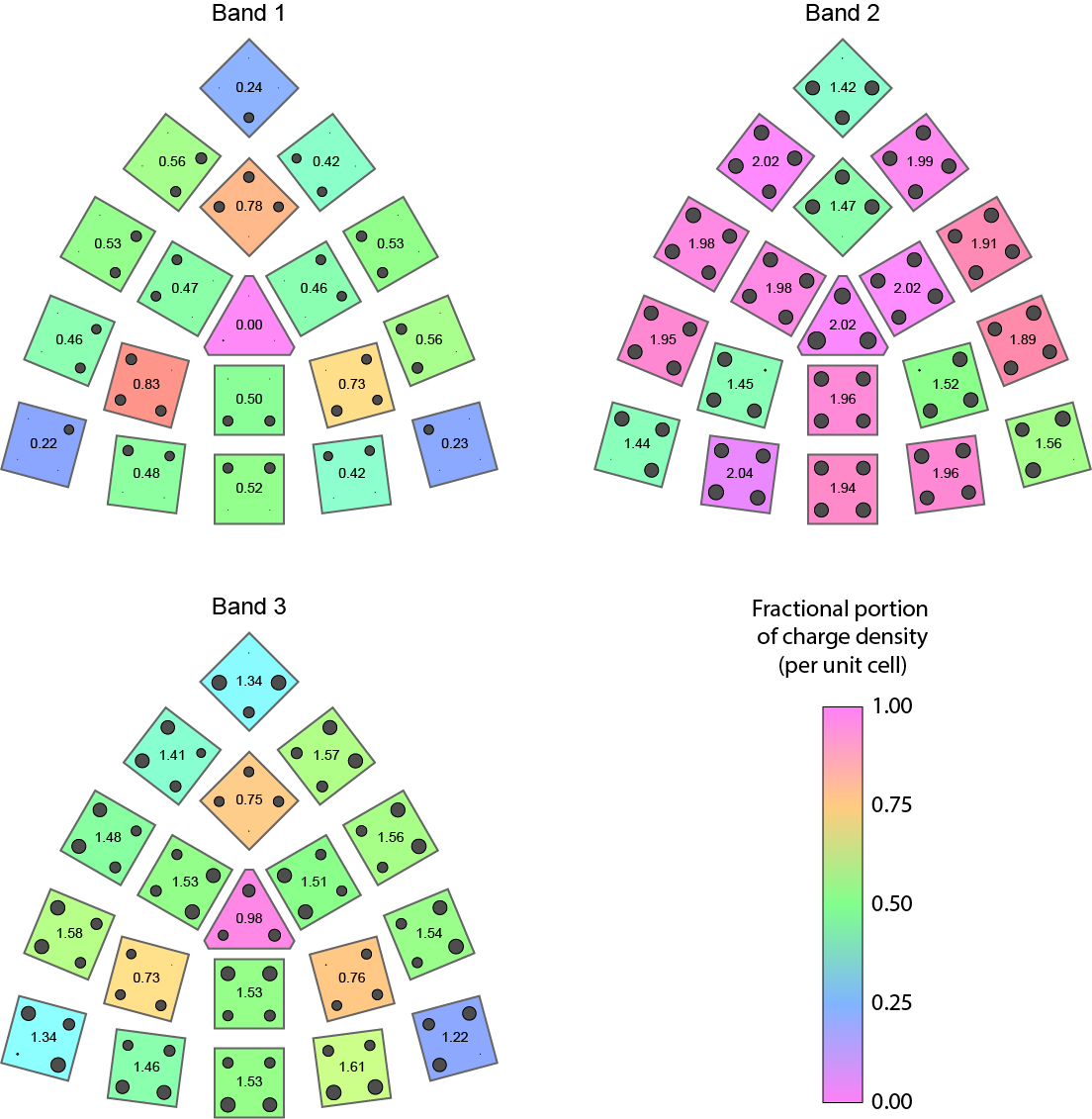}
\caption{Measured fractional mode density for the $C_3$-symmetric insulator after the central unit cell is removed, as shown in Fig. \ref{fig3}(c). The total mode density for each band is listed numerically in each unit cell, and the area of each dot is proportional to the mode density of the corresponding resonator. The color of each unit cell indicates the fractional portion of mode density.}
\label{figs5}
\end{adjustwidth}
\end{figure}

\begin{figure}[h!]
\begin{adjustwidth}{-1in}{-1in}
\centering
\includegraphics[width=\linewidth]{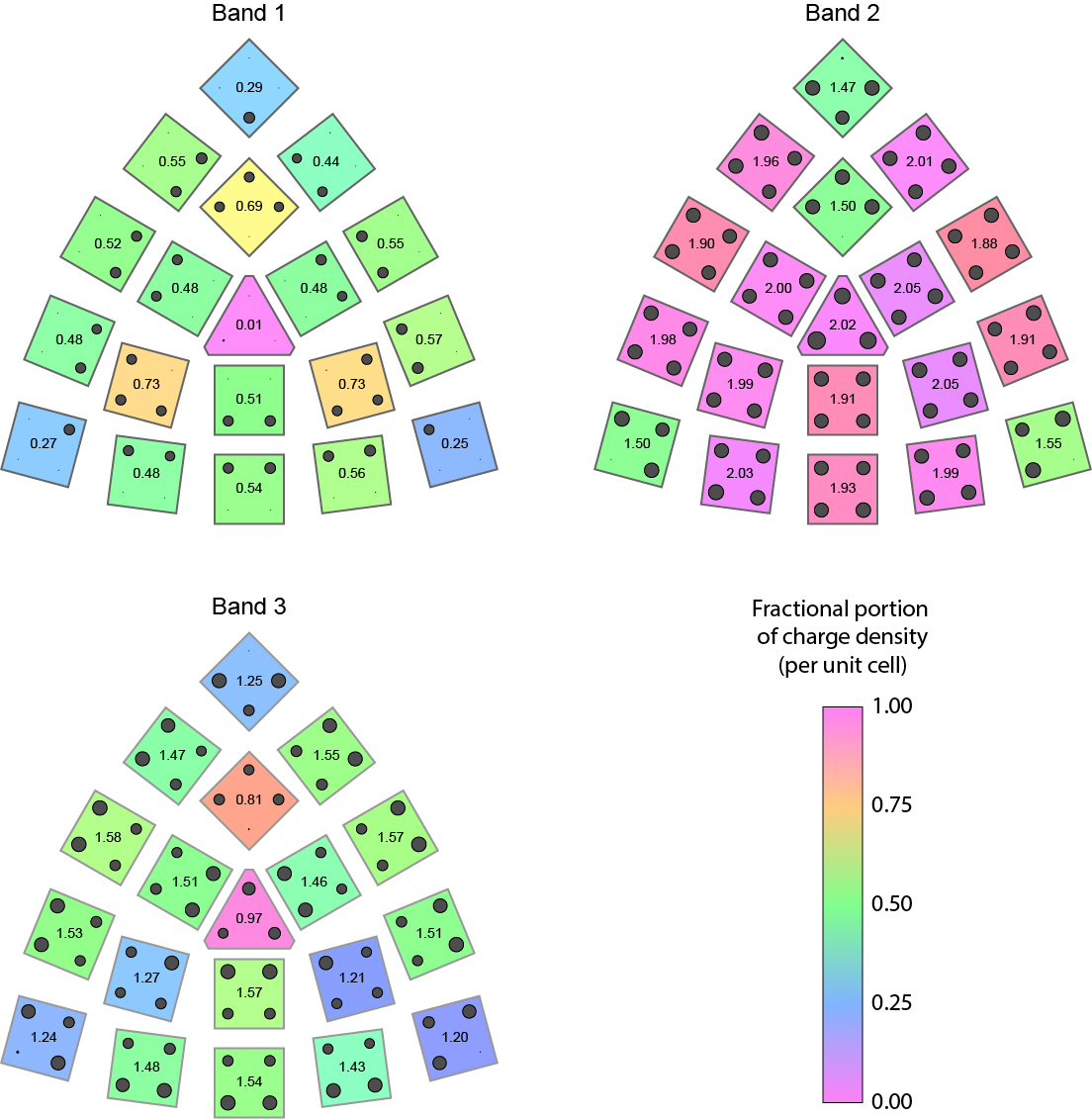}
\caption{Continuing from Fig. \ref{figs5}, now with broken rotation symmetry as in Fig. \ref{fig3}(f). The total mode density for each band is listed numerically in each unit cell, and the area of each dot is proportional to the mode density of the corresponding resonator. The color of each unit cell indicates the fractional portion of mode density.}
\label{figs6}
\end{adjustwidth}
\end{figure}

\begin{figure}[h!]
\begin{adjustwidth}{-1in}{-1in}
\centering
\includegraphics[width=\linewidth]{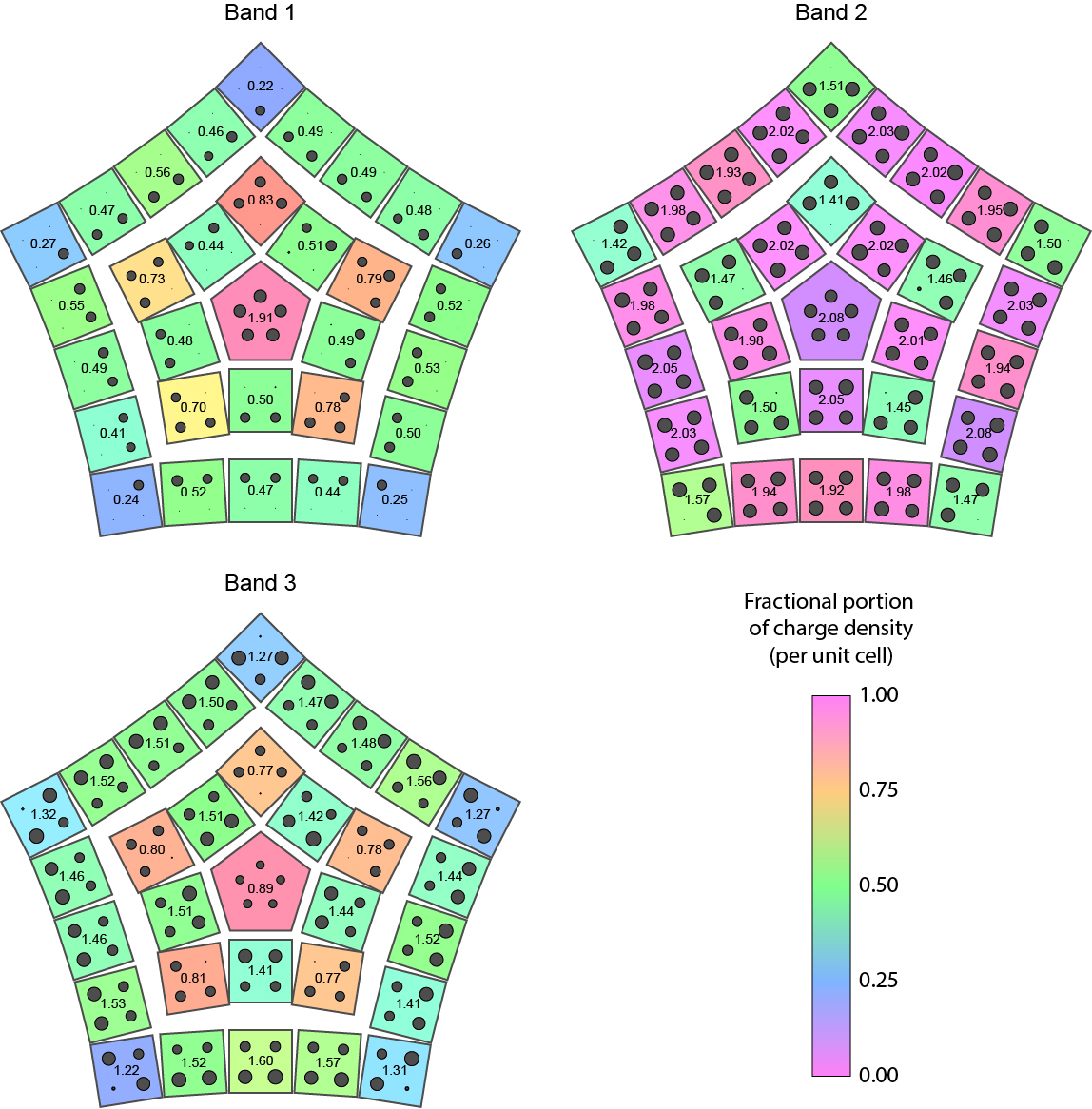}
\caption{Measured fractional mode density for the $C_5$-symmetric insulator after the central unit cell is removed, as shown in Fig. \ref{fig4}(c). The total mode density for each band is listed numerically in each unit cell, and the area of each dot is proportional to the mode density of the corresponding resonator. The color of each unit cell indicates the fractional portion of mode density.}
\label{figs7}
\end{adjustwidth}
\end{figure}

\begin{figure}[h!]
\begin{adjustwidth}{-1in}{-1in}
\centering
\includegraphics[width=\linewidth]{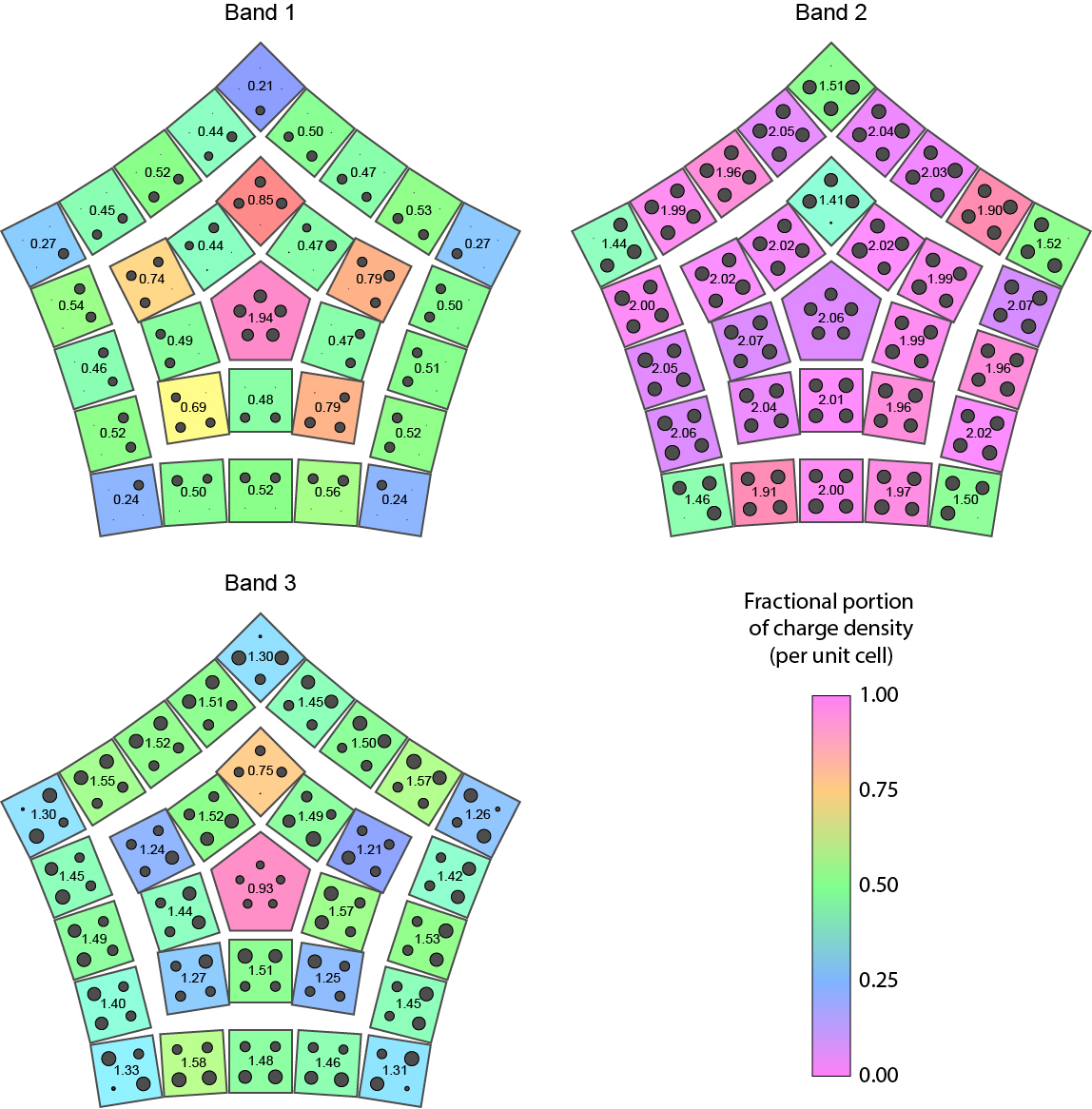}
\caption{Continuing from Fig. \ref{figs7}, now with broken rotation symmetry as in Fig. \ref{fig4}(f). The total mode density for each band is listed numerically in each unit cell, and the area of each dot is proportional to the mode density of the corresponding resonator. The color of each unit cell indicates the fractional portion of mode density.}
\label{figs8}
\end{adjustwidth}
\end{figure}

\end{document}